\documentclass[ aip, PoF,
amsmath,amssymb,
reprint,%
]{revtex4-2}

\usepackage{dsfont}
\usepackage{mathtext}
\usepackage[utf8]{inputenc}

\usepackage{mathtools}
\usepackage[usenames,dvipsnames]{xcolor}
\usepackage{amsmath}
\usepackage{amssymb,mathrsfs}
\usepackage{graphicx}
\usepackage{epstopdf}
\usepackage[colorlinks=true]{hyperref}

\usepackage{mathtext}
\usepackage{subcaption}

\usepackage{etoolbox} 
\usepackage{lipsum} 
\usepackage[capitalize]{cleveref}

\makeatletter
\appto{\appendix}{%
	\@ifstar{\def\theequation@prefix{A.}}%
	{}%
}
\makeatother

\newcommand{\om}{\omega}

\newcommand{\RNumb}[1]{\uppercase\expandafter{\romannumeral #1\relax}}

\begin{document}
\title{Analysis of modulated linear wave packets and gap solitons in coupled optical fibers}

\author{D.~V.~Shaykin}
\affiliation{Russian University of Transport (RUT-MIIT), Obrazcova st. 9, Moscow, Russia}
\affiliation{Skolkovo Institute of Science and Technology, Skolkovo, Moscow, 143026, Russia}
\affiliation{Moscow Institute of Physics and Technology, Institutsky lane 9, Dolgoprudny, Moscow region, 141700, Russia}

\begin{abstract}

A coupled pair of waveguides with refractive indices of different signs is considered. It is shown that the gap in the linear wave spectrum closes if these waves are amplitude modulated. To look for solitons as envelopes of exponentially modulated wave packets, the dispersion law of these packets is used. Different decompositions of the dispersion law lead to solutions related to different parts of the spectrum. Numerical calculations demonstrate a good agreement with analytical predictions.

\end{abstract}

\maketitle

\section{Introduction}

In recent decades, significant attention has been paid to the creation of metamaterials with various properties. In this article, we will focus on a pair of waveguides, one of which has a negative refractive index $n<0$. An important feature of such a material is a negative group velocity 
\begin{equation}
v_g = \frac{\partial \om}{\partial k} \frac{\vec{k}}{k},
\end{equation} 
where $\partial \om / \partial k < 0$. Mandelshtam\cite{{Man1945},{Man1950}} pointed out this connection between refractive index and group velocity in 1940s. Theoretical investigation of propagation of light in the medium started in 1950s\cite{{Siv1957},{Paf1956},{Paf1957},{Paf1959}} and the first experimental samples was obtained only in 21st century\cite{{M1},{M2},{M3},{M4},{M5}}. The creation of the metamaterials resulted in a new wave of theoretical studies aimed at the mediums with a negative refractive index (see, for example, Refs.\cite{{N1},{N2},{N3},{N4},{N5},{N6}}). We are interested in coupled optical fibers with opposite signs of $n$ (see Refs\cite{{C1},{C2},{C3}}). The coupling of modes in waveguides is caused by light tunneling\cite{{T1},{T2},{T3}}.

Following article\cite{Maim2008}, we introduce a system of coupled modes in the form 
\begin{align}\label{main}
&i(\partial A/ \partial t + \partial A /\partial x) -\delta A + p*B + r_1 |A|^2A = 0, \\
&i(\partial B /\partial t - \partial B /\partial x) -\delta B + p*A + r_2 |B|^2B = 0,
\end{align}
where $\delta$ is responsible for the violation of phase-matching and $r_i$ is a nonlinearity parameter. We left the optional parameter $p$, which will be useful for numerical calculations and visualization. The thing is, by dividing this system of equations by $p$, we can rescale the system to arrive at that form. We will use this scaling trick to build a meaningful numerical picture. One can note that \eqref{main} is presented in a dimensionless form; therefore, parameters as well are dimensionless. 

The feature of such a system is a gap in the dispersion law:
\begin{equation}\label{12}
\om = \delta \pm \sqrt{k^2+p^2},
\end{equation}
where the gap of width $2p$ is centered with respect to $\delta$. Generally speaking, a gap in the spectrum is a common occurrence for periodic systems or systems with certain characteristic frequencies; however, our task does not imply anything of the sort.

Recently, it has become\cite{{Int1},{Int2},{Int3},{Int4}} apparent that many nonlinear problems for integrable and non-integrable equations can be solved by studying linear packets. Here, we will show that by investigating a modulated linear packet, we can obtain soliton envelopes in various parts of the spectrum. Previously, similar problems for Bragg waveguides\cite{{I1},{I2},{I3}} were solved using a consequence of a complex method suitable for exceptional cases of integrable or nearly integrable equations. In our task, we will not rely on such a strong and limiting features, but will follow the method\cite{{A1},{A2}} suitable for any equations, as, for example, it was done in the  Refs\cite{{Kamch1},{Kamch2}}.

\section{Wave packets}

Let us look for a solution in the form of a linear wave packet with a carrier wave number $k$ and a frequency $\om$
\begin{align}\label{21}
&A(x,t) = A(\xi)e^{i(kx-\om t +\phi(\xi))}, \\
&B(x,t) = \big[ B(\xi)+ib(\xi)\big]e^{i(kx-\om t + \phi(\xi))},
\end{align}
where the variable $\xi=x-Vt$ is related to the motion of the center of mass of the packet with velocity $V$, $\phi$ is a chirp function. Substitution this form into \eqref{main} gives
\begin{equation}\label{22}
\begin{split}
A\big[\om_0-k+(V-1)\dot{\phi}\big]+pB+r_1A^3&=0,
\\
\dot{A}(V-1) -pb&= 0,\\
B\big[ \om_0+k+(V+1)\dot{\phi}\big]+\dot{b}(V+1)&+\\
+pA+r_2(B^2+b^2)B&=0,\\
b\big[ \om_0+k+(V+1)\dot{\phi}\big] - \dot{B}(V+1)&+\\
+r_2(B^2+b^2)b&=0.
\end{split}
\end{equation}
Such a system describes an evolution of the general view of wave packet \eqref{21}. Let us specify it by modulating the amplitude functions by $exp(-\Omega\xi)$, that is
\begin{equation}
\big(A,B,b\big) = \big(A_0,B_0,b_0\big) e^{-\Omega\xi}
\end{equation}
Since we are interested in linear wave packets, then we need to linearize this system:
\begin{equation}\label{23}
\begin{split}
A\big[\om_0-k\big]+pB&=0,
\\
\dot{A}\Omega(V-1) +pb&= 0,\\
B\big[ \om_0+k\big]-\dot{b}\Omega(V+1)+pA&=0,\\
b\big[ \om_0+k\big] + \dot{B}\Omega(V+1)&=0,
\end{split}
\end{equation}
where we introduce a new convenient parameter $\om_0 = w-\delta$. As we can see there isn't function $\phi$. Therefore, we have $4$ equations and only $3$ parameters $\om,\Omega,k$, so after simplification we arrive at the two equations
\begin{equation}\label{24}
V = \frac{k}{\om_0},\qquad k^2 = \frac{\om_0^2+\Omega^2-p^2}{1+\Omega^2/\om_0^2} 
\end{equation}
They have simple physical meanings, its can be figured out from the limit $\Omega \to \infty$ (the case of unmodulated waves); the first one denotes a dispersion law for the modulated waves, and the second one is the group velocity. As one can see from such a dispersion law, the gap at $k=0$ narrows, and while $\Omega>0$ it disappears at all.
So it allows us to look for the wave packets that have the carrier frequency located at the forbidden earlier zone; the soliton as an envelope of such packets is called a gap soliton. Generally speaking since we investigate the wave packets, we should  take into account $\Omega \ll \om$, so we introduce a small parameter
\begin{equation}
\varepsilon = \frac{\Omega}{\om} 
\ll 1.
\end{equation}
It means the envelopes of wave packets should be much slower functions than wave packet oscillations, and to investigate an evolution of the envelope, we need to use the slow variable $\Omega \xi$. Then for the envelope $\partial / \partial \xi = \Omega \partial / \partial \xi = \varepsilon \omega \partial /\partial \xi$, and general system \eqref{22} transforms to
\begin{equation}\label{25}
\begin{split}
A\big[\om_0-k+(V-1)\dot{\phi}
\Omega\big]+pB+r_1A^3&=0,
\\
\dot{A}
\Omega(V-1) -pb&= 0,\\
B\big[ \om_0+k+(V+1)\dot{\phi}
\Omega\big]+\dot{b}\Omega(V+1)&+\\
+pA+r_2(B^2+b^2)B&=0,\\
b\big[ \om_0+k+(V+1)\dot{\phi}\Omega\big] - \dot{B}\Omega(V+1)&+\\
+r_2(B^2+b^2)b&=0.
\end{split}
\end{equation}

\section{Types of solitons}
For the further work it will be convenient to use the inverse dispersion law in the form
\begin{equation}\label{31}
k = \sqrt{\om_0^2+(\varepsilon \om)^2-p^2}/\sqrt{1+(\varepsilon \om)^2/\om_0^2}.
\end{equation}
As one can note, we can decompose such an expression in different ways, and it leads to different solutions.

\subsection{Outside the gap}

\begin{figure}
\begin{center}
	\includegraphics[width = 7cm,height = 7cm]{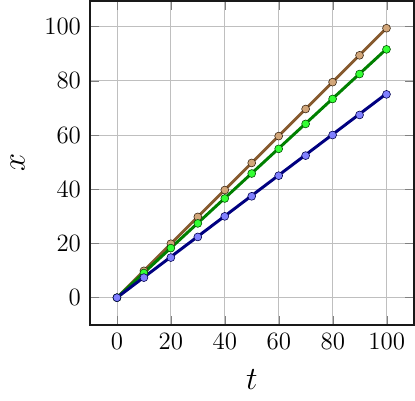}
\caption{Comparison of the numerical solution (circles) of equation \eqref{main} with the analytical predictions (solid lines) of equation \eqref{24}for the solitons outside the gap for the following three datasets $\{\om,\delta,p,\varepsilon\}$: brown $\{12,2,1,0.2 \}$, green$\{15,10,2,0.05\}$, blue $\{5,2,2,0.1\}$; $r_2=r_1 = 1$.}
\label{fig1}
\end{center}
\end{figure}

For the beginning, we obtain an ordinary soliton outside the gap, that is $\om_0>p$. Therefore we can decompose \eqref{31} as
\begin{equation}\label{32}
k = \sqrt{\om_0^2-p^2}+\frac{(\om p/\om_0)^2}{2\sqrt{\om_0^2-p^2}}\varepsilon^2+o(\varepsilon^2) = \alpha+\gamma\varepsilon^2+o(\varepsilon^2).
\end{equation}
After substitution \eqref{32} into \eqref{25} we can look for a solution in series form $f = \sum_0^\infty \varepsilon^n f_n$. Simple calculations lead to the following 
\begin{equation}\label{33}
\{A,B,\dot{\phi}\} = \sum_{n=1}^{\infty}\varepsilon^n\{A_n,B_n,\dot{\phi_n}\},\qquad b = \sum_{n=2}^\infty \varepsilon^n b_n  
\end{equation} 
In the first order Eqs.\eqref{25} give
\begin{equation}
A_1(\om_0-\alpha)+pB_1 = 0,
\end{equation}
the second order leads to
\begin{equation}
\begin{split}
A_2(\om_0 -\alpha)+pB_2 = 0,\\
\dot{A_1}\om\big(\frac{\alpha}{\om_0}-1\big) = pb_2.
\end{split}
\end{equation}
Finally, in the third order we obtain the well-known oscillatory equation 
\begin{equation}\label{os1}
\ddot{A_1} = A_1- \frac{\om_0^2}{\om^2}\Big[ \frac{r_2(\om_0-\alpha)^3+r_1p^2(\om_0+\alpha)}{p^3}\Big]A_1^3
\end{equation}
that has a soliton solution. Thus we have in the leading order 
\begin{equation}
A = \frac{\varepsilon e^{i(kx-\om t)}}{\cosh{(\varepsilon \om \xi)}} \frac{\sqrt{2p^3}}{\sqrt{r_2(\om_0-\alpha)^3+r_1p^2(\om_0+\alpha)}}\frac{\om}{\om_0},
\end{equation}
and $B = -A(\om_0-\alpha)/p$. In particular, if $p\ll \om_0 $, then $|B|\ll|A|$, that is, we come to the situation of coupled waveguides in which the wave propagates along only one of them.

As can be seen from the Eq.\eqref{os1} the solitons don't exist for a pair of defocusing waveguides ($r_2, r_1 <0$).

\begin{figure}
\begin{center}
	\includegraphics[width = 7cm,height = 7cm]{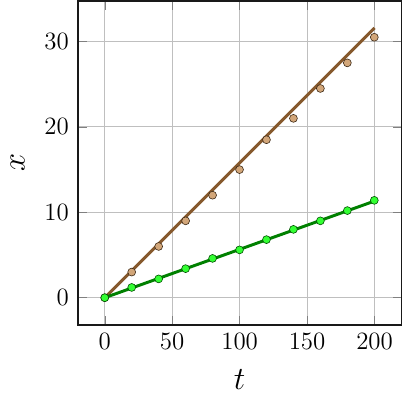}
\caption{Comparison of the numerical solution (circles) of equation \eqref{main} with the analytical predictions (solid lines) of equation \eqref{24}for the solitons on the gap for the following two datasets $\{\delta,p,\varepsilon\}$: brown $\{15,1,0.01 \}$, green$\{ 2,15,0.05\}$; $r_2=r_1 =1$.}
\label{fig2}
\end{center}
\end{figure}

\subsection{On the gap}
Another decomposition of Eq.\eqref{31} we can obtain if we take $\om_0 = p$:
\begin{equation}
k = \om \varepsilon + o(\varepsilon^2).
\end{equation}
In this situation we can as well use the series \eqref{33}. The first order gives
\begin{equation}
A_1 + B_1 = 0,
\end{equation}
the second order gives
\begin{equation}
p(A_2+B_2) = A_1 \om.
\end{equation}
On the third step we obtain
\begin{equation}\label{os2}
\ddot{A_1} = A_1 - (r_2+r_1)\frac{p}{\om^2}A_1^3,
\end{equation}
hence in the leading order we have
\begin{equation}
A = \varepsilon\frac{\sqrt{2}\om}{\sqrt{(r_2+r_1)p}}\frac{e^{i(kx-\om t)}}{\cosh{(\varepsilon \om \xi)}},
\end{equation}
and $B = -A$.

\begin{figure}
\begin{center}
	\includegraphics[width = 7cm,height = 7cm]{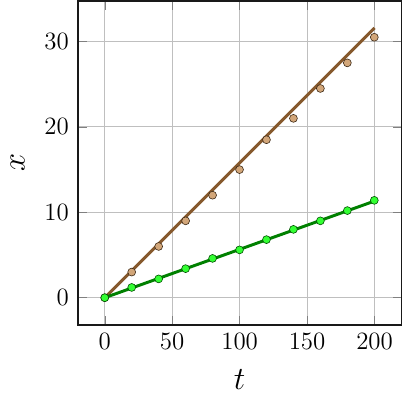}
\caption{Comparison of the numerical solution (circles) of equation \eqref{main} with the analytical predictions (solid lines) of equation \eqref{24}for the solitons inside the gap for the following two datasets $\{\delta,p,d,\varepsilon\}$: brown $\{0.1,1,0.4,0.1 \}$, green$\{ 10,1,20,0.01\}$;$r_2=r_1 =1$.}
\label{fig3}
\end{center}
\end{figure}

\subsection{Inside the gap}

To decompose the dispersion law \eqref{31}, we have to demand $\om_0^2-p^2+(\varepsilon \om)^2>0$. We can meet this requirement as well inside the gap, in particular, if we take $w_0 = p-q\varepsilon^2$, where parameter $q$ has a limitation that can be obtained by substituting it into $\om_0^2-p^2+(\varepsilon \om)^2>0$. A simple analysis shows $q\lesssim (\delta+p)^2/2p$ by implying $\om \sim \delta$. Another limitation comes from the denominator of the formula \eqref{31}: $\varepsilon\ll \om_0/\om$. In such a framework we can decompose Eq.\eqref{31} near the initial gap boundary but inside it:
\begin{equation}
k = \varepsilon \sqrt{(\delta+1)^2-2*qp}+o(\varepsilon^2) = \alpha\varepsilon+o(\varepsilon).
\end{equation}
Series \eqref{33} also satisfy our problem and finally we can arrive at
\begin{equation}\label{os3}
\ddot{A_1} = A_1 - (r_2+r_1)\frac{p}{(\delta+p)^2}A_1^3
\end{equation}
that solves the task:
\begin{equation}
A = \varepsilon\frac{\sqrt{2}(\delta+p)}{\sqrt{(r_2+r_1)p}}\frac{e^{i(kx-\om t)}}{\cosh{(\varepsilon \om \xi)}},
\end{equation} 
and $B = -A$.

One can note, in the last two situations, the wave number $k \propto \varepsilon$ and the envelope function are also proportional to $\varepsilon$. It means that in these cases the wave packets include only a few oscillations. Also, the solitons exist for the different waveguides only if $r_2+r_1>0$.

\section{Conclusion}

Based only on the linear wave packet's theory, we were able to obtain solitons as envelopes in different parts of the spectrum. This idea doesn't depend on the integrability of any equations; as a consequence, it can be applied to the more complicated problems.

\section{Acknowledgment}
I would like to express my gratitude to my teacher A.M. Kamchatnov for his constant help in my work. He told me the method used here.

\section{Sources of funding}

The work was carried out as part of the state assignment dated 20.03.2025 No. 103-00001-25-02 on the topic "Optical properties of dimer chains of laser waveguides with topological defects, including nonlinear ones and those with a negative refractive index."

\end{document}